\documentclass[conference]{IEEEtran}
\usepackage{comment}
\usepackage{multirow}
\usepackage[dvipsnames]{xcolor}
\usepackage[utf8]{inputenc}
\usepackage{booktabs}
\usepackage{comment}
\usepackage{booktabs}
\usepackage{array,ragged2e}
\newcolumntype{P}[
1]{>{\RaggedRight\arraybackslash}p{#1}}
\usepackage{stfloats} 
\usepackage{makecell}

\ifCLASSINFOpdf
  \usepackage[pdftex]{graphicx}
\else
\fi
\usepackage{hyperref} 

\begin{document}
%
\title{Maturity of Vehicle Digital Twins: \\From Monitoring to Enabling Autonomous Driving}
%
%

\author{\IEEEauthorblockN{Robert Klar\IEEEauthorrefmark{1}\IEEEauthorrefmark{2},
Niklas Arvidsson\IEEEauthorrefmark{2}, 
Vangelis Angelakis\IEEEauthorrefmark{1}
}
\IEEEauthorblockA{\IEEEauthorrefmark{1}Department of Science and Technology, Linköping University,
Campus Norrköping, 60 174, Sweden}
\IEEEauthorblockA{\IEEEauthorrefmark{2}Swedish National Road and Transport Research Institute (VTI), SE-581 95 Linköping, Sweden}
E-mail: robert.klar@liu.se, niklas.arvidsson@liu.se, vangelis.angelakis@liu.se
}

\maketitle
\thispagestyle{plain}
\pagestyle{plain}

\begin{abstract} 
Digital twinning of vehicles is an iconic application of digital twins, as the concept of twinning dates back to the twinning of NASA space vehicles. Although digital twins (DTs) in the automotive industry have been recognized for their ability to improve efficiency in design and manufacturing, their potential to enhance land vehicle operation has yet to be fully explored. Most existing DT research on vehicle operations, aside from the existing body of work on autonomous guided vehicles (AGVs), focuses on electrified passenger cars. However, the use and value of twinning varies depending on the goal, whether it is to provide cost-efficient and sustainable freight transport without disruptions, sustainable public transport focused on passenger well-being, or fully autonomous vehicle operation. In this context, DTs are used for a range of applications, from real-time battery health monitoring to enabling fully autonomous vehicle operations. This leads to varying requirements, complexities, and maturities of the implemented DT solutions. This paper analyzes recent trends in DT-driven efficiency gains for freight, public, and autonomous vehicles and discusses their required level of maturity based on a maturity tool. The application of our DT maturity tool reveals that most DTs have reached level 3 and enable real-time monitoring. Additionally, DTs of level 5 already exist in closed environments, allowing for restricted autonomous operation. 
\end{abstract}

%
\IEEEpeerreviewmaketitle

\section{Introduction}
Digital twinning of vehicles is an iconic application of digital twins, as the idea of twinning dates back to NASA's Apollo program, where at least two identical space vehicles were built to mirror the conditions of the vehicle in space and to provide decision-support during the mission based on simulations using the vehicle's twin on Earth \cite{rosen2015importance}.

While digital twins are still of significance in the context of vehicles in the aerospace industry \cite{yang2021application}, their use is constantly expanding to new vehicles. These include vehicles in freight \cite{tran2023advanced} and passenger transportation \cite{wang2022mobility}, as well as autonomous vehicles for various Industry 4.0 purposes, such as last-mile delivery \cite{DT_last_mile}, container handling in ports \cite{yang2022digital} and automated guided vehicles (AGVs) \cite{lichtenstern2022data}.

A key development here is that vehicles are becoming software-driven objects, requiring a large number of electronic control units (ECUs) that communicate over complex in-vehicle networks. This evolution is driven by the industry's need to develop new automotive functions, such as adaptive cruise control \cite{popa2022cartwin}.

The continuous expansion of digital twin applications to new vehicle types has the potential to enable the exchange and interconnection of digital twins of different vehicle types and their surrounding environments. However, there is also the risk that digital twins of different vehicle types will take different paths in terms of understanding, functionality and implementation of digital twins. 

The purpose of this paper is to address this gap by presenting the current state of digital twinning for three types of vehicles: freight, passenger, and autonomous vehicles, and evaluating their level of maturity using a maturity tool \cite{klar2023digital}. 

\section{Digital Twins}

\begin{figure}
\centering
\includegraphics[width=1.0\linewidth]{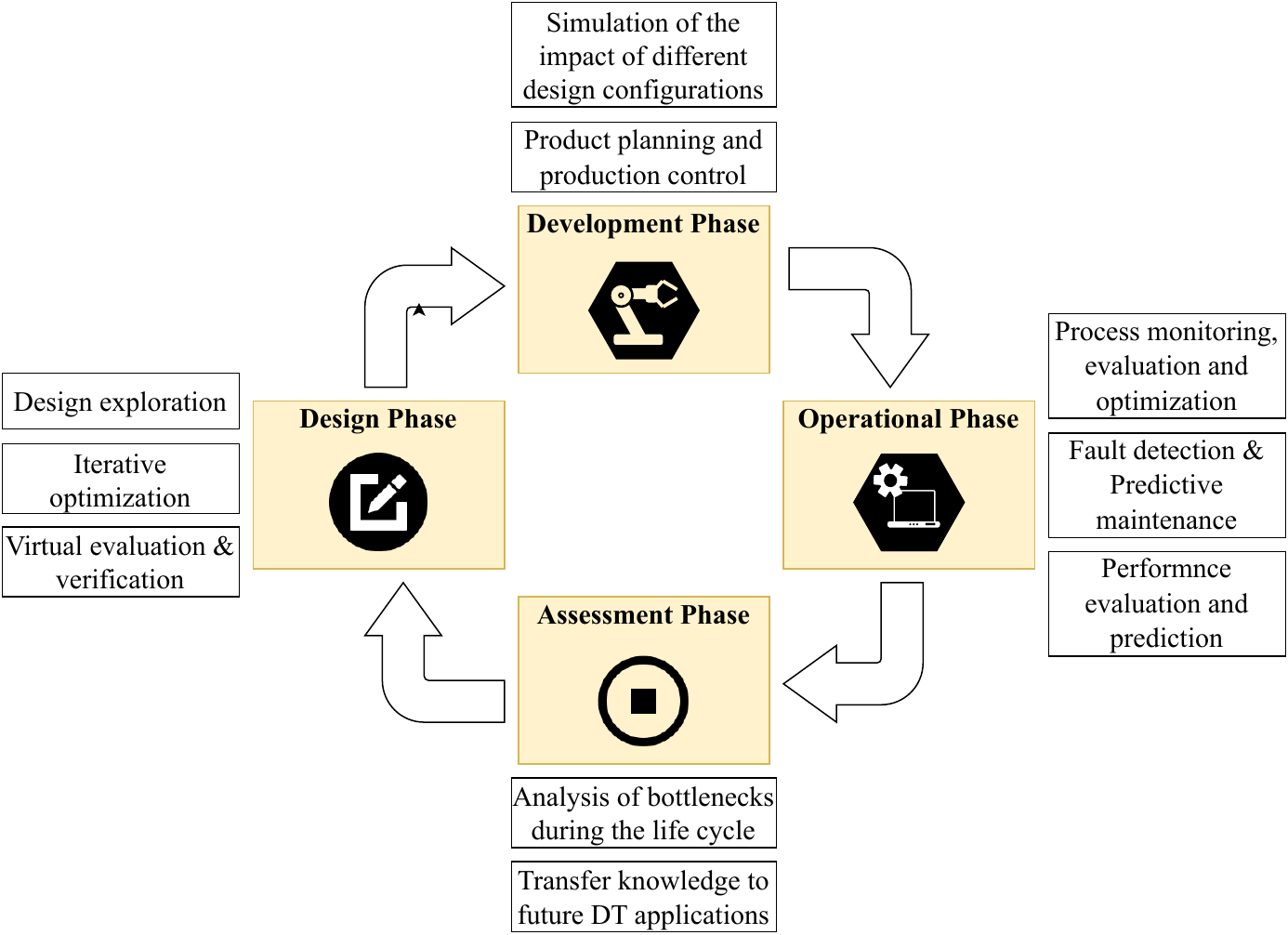}
\caption{Applications of DTs throughout their lifecycle. Adapted from \cite{liu2021review}.}
\end{figure}\label{Figure:DT_lifecycle}

\subsection{Characteristics of Digital Twins}\label{section:DT_characteristics}
The Digital Twin Consortium defines a digital twin (DT) as ``a virtual representation of real-world entities and processes, synchronized at a specified frequency and fidelity''\cite{Olcott2020digital}. According to Piromalis and Kantaros in paper \cite{piromalis2022digital}, digital twins are designed on a product, process or systems (including system of systems) level. 
According to Grieves, who coined the concept of the Digital Twin, initially termed as Product Lifecycle Management, there are three different types of digital twins, depending on the stage of development of the product and its digital twin \cite{grieves2023digital}. 
The three different digital twin types are presented in Table \ref{Table:DigitalTwinTypes}. The Digital Twin Prototype (DTP) allows for testing new components under various conditions to implement design improvements and predict vehicle performance. It thus enables determination of the optimal configuration. Once the vehicle is manufactured, it receives its own Digital Twin Instance (DTI), which mirrors and optimizes the vehicle based on real-time data. Finally, various DTIs form a Digital Twin Aggregate (DTA), which allows for the exchange of data and knowledge and joint optimization among multiple vehicles, as in the case of platooning. 

The four phases within a product lifecycle during which the product is optimized by a DT, and the resulting potential efficiencies enabled by digital twins, are illustrated in Figure \ref{Figure:DT_lifecycle}. During the design and development phase, the DTP is the prominent DT type, while the DTI is used during the operational phase to improve efficiency. Multiple DTIs are aggregated to DTAs, allowing for learning from the twinning experiences of several instances and transferring their knowledge to upcoming vehicle twins during their assessment phase.

Digital twins are thus implemented to perform lifecycle-specific simulation tasks, closing the loop from operations and service back to the design phase \cite{spiryagin2023vehicle}. In the context of smart cities, transportation, and logistics, a narrow interpretation of digital twins may fall short. Beyond technology, these domains involve active participation from diverse communities and stakeholders \cite{mylonas2021digital}. Leveraging data analytics, AI, machine learning (ML), and agent-based modeling, digital twins may address critical what-if scenarios \cite{rizopoulosdigital}. 

\begin{figure}
\centering
\includegraphics[width=1.0\linewidth]{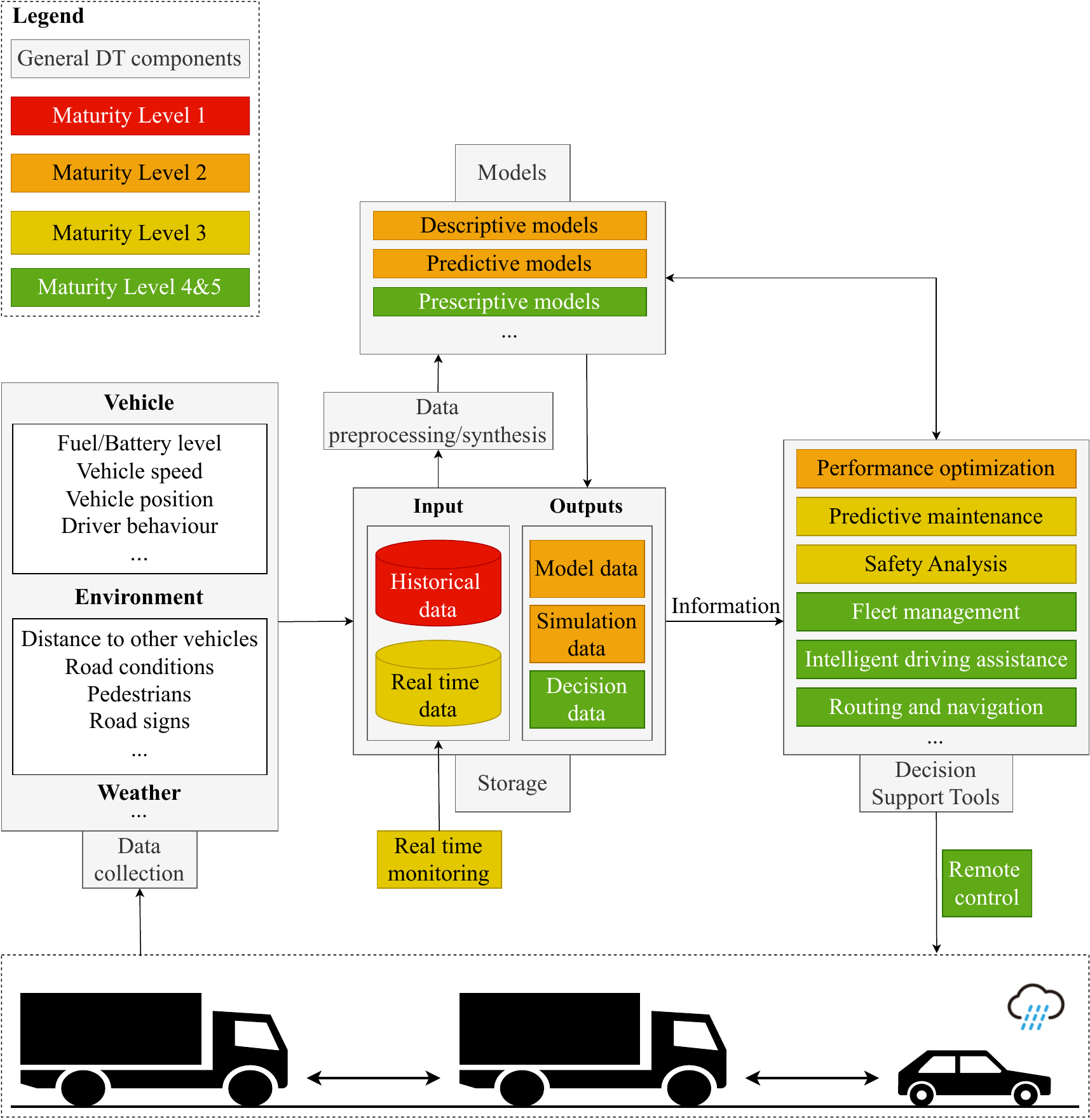}
\caption{Application of the maturity levels in the DT/reality context.}\label{Figure:vehicle_DT}
\end{figure}

Figure \ref{Figure:vehicle_DT} illustrates the various components of a digital twin, including data collection, modeling, and decision support tools that enable DT-driven vehicle management and operation. The descriptive models of the digital twin facilitate understanding of current and past events, while the predictive models allow prediction of information that hasn't yet been measured, enabling decision making and trade-off analysis \cite{eramo2021conceptualizing}. Finally, prescriptive models allow the vehicle to be controlled based on the results of the predictive models. Different DT characteristics and capabilities are highlighted according to the maturity levels outlined in section \ref{section:DT_maturity}.

\begin{table*}
    \caption{Digital twin types (adapted from \cite{grieves2023digital}) and their potential in the vehicle context}
    \centering
    \renewcommand{\arraystretch}{1.3}
    \begin{tabular}{ P{4.0cm} p{13cm}}
        \toprule
        \textbf{Digital Twin Type} & \textbf{Characterization and gained value}\\
        \midrule
        Digital Twin Prototype (DTP) & The Digital Twin Prototype is created prior to assembling the actual product. It is used to test and validate the best vehicle configuration based on operational data from previous vehicle generations and their respective twins (if available). The goal is to achieve the most efficient vehicle setup while minimizing physical testing. \\
        Digital Twin Instance (DTI) & The DTP transitions into a unique Digital Twin Instance (DTI) for each manufactured vehicle. It enables monitoring, optimization, and partial remote control of the vehicle through bi-directional data exchange. The corresponding DTI develops into a self-learning system through the accumulation of data from the vehicle and related instances, given a high level of interoperability.\\
        Digital Twin Aggregate (DTA) & Different DTIs are then aggregated into a Digital Twin Aggregate (DTA). This enables the synthesis of data and knowledge obtained from different DTIs to assess actions and resulting patterns among other DTIs. It also improves the efficiency of each future vehicle and its respective DTI by learning from past experiences.                Additionally, it enables many multi-vehicle joint operations, such as platooning.  \\
        \bottomrule
    \end{tabular}
    \label{Table:DigitalTwinTypes}
\end{table*}

\subsection{Digital twins in the context of vehicles}\label{section:vehicles}
DTs are used in the design and production of new vehicles, as well as in measuring specific patterns and functional information of the vehicle after manufacturing to aid in performance improvement \cite{piromalis2022digital}.
Digital twins are thus considered a valuable tool for optimizing the efficiency and reliability of electric vehicles, from design to cost-effective operation \cite{van2021beyond}. 

Van Mierlo et al. state in paper \cite{van2021beyond} that digital twins in the context of vehicles are enabled by the ongoing development of five complementary technologies: the Internet of Things (IoT), cloud computing, APIs and open standards, artificial intelligence (AI), and digital reality technologies.

Vehicles are experiencing a constant increase in the number of sensors and electronic control units. This growth is coupled with the development of safety systems, including advanced driver assistance systems (ADAS) and automated driving systems (ADS). These systems require vehicles to be aware of their surroundings, make decisions, and take action, making them ideal candidates for the deployment of DTs \cite{schwarz2022role}. 

Recent research indicates that digital twins should not operate in isolation, but rather engage in constant communication to improve situational awareness beyond the twin's boundaries and facilitate joint decision-making \cite{klar2023digital}. In this context, three types of communication around digital twins have been identified \cite{schwarz2022role}. These communication channels include bi-directional communication between the digital and physical twin, communication among multiple connected digital twins, and communication between digital twins and domain experts through user interfaces \cite{barricelli2019survey}. The coordination of digital twins for joint decision-making is crucial for various applications, such as fleet management or DT-guided platooning. This requires interoperability among digital twins, which is identified as the highest level of maturity in section \ref{section:DT_maturity}.

\subsection{Digital Twins - A Maturity Perspective}\label{section:DT_maturity}
\begin{figure*}
\centering
\includegraphics[width=1.0\linewidth]{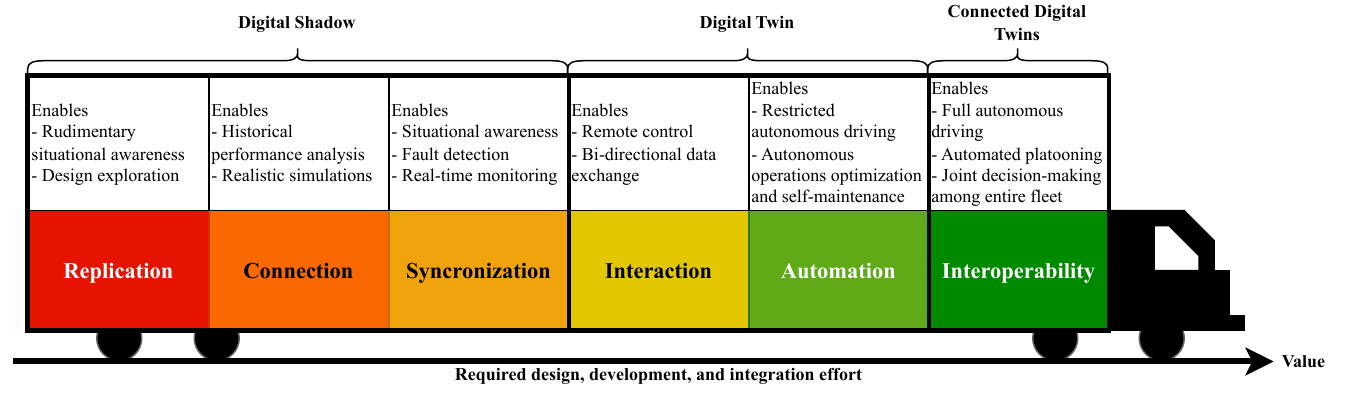}
\vspace*{-0.8cm}
\caption{Maturity levels and their main enabling potential in the vehicle context. Adapted from \cite{klar2023digital}.}\label{Figure:maturity levels}
\end{figure*}

Digital twins focus on dynamics, learning, and evolution, rather than merely being digital representations of static real-world objects \cite{klar_DT}. There are six maturity levels that encompass the maturity of digital twins spanning from the acquisition of physical system properties to interoperability between digital twins \cite{klar2023digital}. These six maturity levels are adapted from \cite{klar2023digital} and can be summarized as follows:

\begin{enumerate}
    \item \textbf{Replication of assets:} Enables an exact replication of the vehicle to be twinned. Particularly relevant for DTPs during the design and development phase.
    \item \textbf{Connection of models and systems:} Enables capturing and simulating the effects of real-world events based on historical data. It allows for testing configurations during the design phase and what-if simulations in the operational phase. The following levels are most relevant to the operational phase.

    \item \textbf{Synchronization of data and processes:} Enables the real-time detection and simulation of the impact of changes, such as traffic disruptions, through timely mapping from the physical world to the digital world. Allows a user to remotely monitor an unmanned vehicle. 
    \item \textbf{Interaction between the Digital Twin and the Asset:} Enables bidirectional exchange of information, allowing the user to issue control commands remotely. Allows a user to remotely control an unmanned vehicle.
    \item \textbf{Automation of processes:} Enables a vehicle's DT to make autonomous decisions about operations and maintenance based on real-time and historical data and its models. Reaching this level enables autonomous vehicles.
    \item \textbf{Interoperability across Digital Twins:} Enables the optimization of a vehicle beyond its physical boundaries by exchanging information and knowledge with other related DTIs. This allows for joint decision-making with DTIs of other vehicles and other DTs within their environment. This enables the full potential of DTAs.
\end{enumerate}

Figure \ref{Figure:maturity levels} illustrates the different maturity levels and their key enabling potential, and also provides a mapping of these levels to the digital twin constructs of digital shadow \cite{botin2022digital}, digital twin, and connected digital twins \cite{lamb2022gemini}. It demonstrates that digital twins can be implemented in consecutive stages and be expanded depending on their respective objective.

\section{Digital Twins of Vehicles - potential and maturity}

\begin{figure}
\centering
\includegraphics[width=0.9\linewidth]{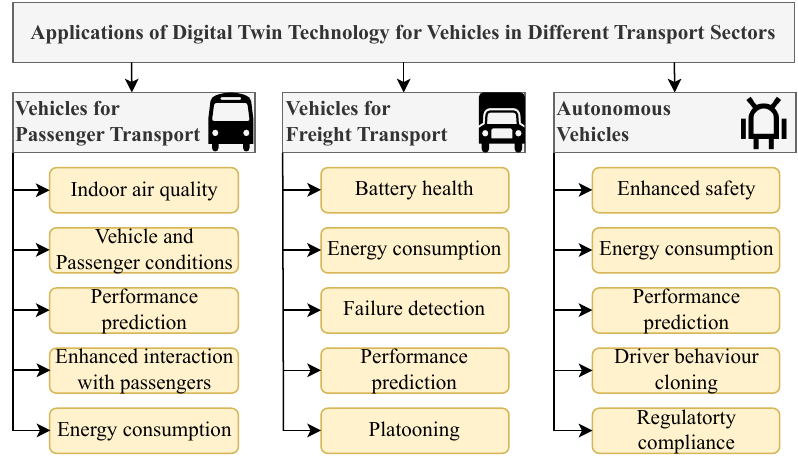}
\label{Figure:DT_applications}
\vspace*{-0.3cm}
\caption{Application of digital twins for different types of vehicles.}\label{DT_vehicle_applications}
\end{figure}

Digital twins have become increasingly important for various types of vehicles due to their ability to enhance efficiency and safety, as well as enable autonomous driving \cite{schwarz2022role}. These vehicles include those for public transportation \cite{botin2021digital}, where passenger comfort is a top priority, freight transportation \cite{tran2023advanced}, where the focus is on cost-effective and environmentally friendly transportation of goods, and autonomous vehicles. Recent advancements in closed environments such as ports \cite{yang2022digital}, warehouses \cite{barosan2020development}, and mining \cite{ai2023pmworld} have outpaced those in open environments like public road transport \cite{bednarz2024framework} or last mile logistics \cite{schnieder2024digital}. 
An overview of applications of digital twins for passenger transport vehicles, freight transport vehicles, and autonomous vehicles is presented in Figure \ref{DT_vehicle_applications} 
The following subchapters present recent contributions for these vehicle types and their required level of maturity.

\subsection{Vehicles for freight transport}
Digital twins are considered a means to improve reliability and ensure smooth delivery processes via truck \cite{tang2023deterioration}. In paper \cite{tang2023deterioration}, the authors present a framework for constructing digital twins of a truck fleet. The purpose of this framework is to create a virtual space that can be used to calculate the most efficient inspection plan. The authors utilized data analytics to identify trends related to safety component failures, such as brakes, tires, or lights, by analyzing historical. This allows for the DT-based detection of failures in real-time data based on the patterns obtained from the historical data.

Moreover, digital twins are considered an enabler for the electrification of freight trucks and other vehicles. This is an important goal since transportation is responsible for over 18\% of global carbon dioxide emissions\cite{bhatti2021towards}.  
In order to electrify vehicles on a large scale and improve their operational performance, it is necessary to develop supporting architecture that can optimize them in a sustainable manner. In this context, Digital Twins are perceived as the enabling tool to achieve this \cite{bhatti2021towards}.
Such a digital twin framework and digital twin platform is proposed by Tran et al. in \cite{tran2023advanced}. It consists of a universal DT architecture that allows for different DT models covering different applications such as monitoring and prediction of energy consumption, battery aging characterization, and monitoring of cooling circuits and thermal system.

Digital twins are also considered a tool to enable platooning. Truck platooning has the potential to improve safety, security, and efficiency, and control overall traffic flow while reducing resource usage \cite{javed2021safe}. In  paper \cite{javed2021safe}, the authors proposed a simulation-based digital twin for verifying and validating as well as fine-tuning the platooning strategy.

The three cases presented here all occur during the operational phase. Although most trucks and their corresponding DTs are considered instances, their applications benefit from accumulating data and knowledge from the entire DTA that represents the fleet of trucks. To enable DT-based inspection planning and DT-based monitoring of operational conditions, a DT of level 3 is required. If the vehicle is controlled remotely by the user of the twin or by the twin itself to perform maintenance actions, a maturity level of 4 or 5 is required. In the case of platooning, maturity level 6 is required, as the different DTs representing the truck platoon are in constant exchange to make the best decision for the entire fleet.

\subsection{Vehicles for passenger transport}
Similar to freight vehicles, digital twins are also recognized as a tool to improve the efficiency of passenger vehicles, as well as the comfort and safety of its passengers. The use of a digital twin for a bus enables the monitoring and prediction of its dynamic performance, analysis of factors affecting energy usage, and monitoring of current and future conditions of the vehicle (such as the battery) and its passengers (including indoor air quality) \cite{ruiz2022digital}.

Bot{\'\i}n-Sanabria et al. present the application of a digital twin for a bus with various capabilities. These capabilities include performance monitoring, performance prediction based on operational changes and its interaction with the environment, and information sharing with the driver and other users \cite{botin2021digital}.

Beyond the goal of improving the efficiency of individual vehicles, digital twins are being applied to entire urban transportation systems. Digital twins are thus seen as an enabling technology for effective and sustainable traffic management in road networks. With intelligent traffic monitoring methods, information can be transferred between vehicles and roads. As a result, traffic congestion can be reduced, the capacity of the road network increased, the number of accidents reduced, and energy and pollution savings achieved \cite{utku2023digital}.

Effective joint optimization of the entire road network and the vehicles operating on it requires joint decision making between the road network and its vehicles. This requires a high level of integration between the DTs or information systems representing the vehicles and the road network, interoperability, standardization and a common data environment \cite{broo2023digital}.

The three cases presented demonstrate that DTs have significant potential to increase the efficiency of the vehicle and the comfort of its passenger, but also of the entire transportation system. By monitoring the condition of the vehicle, both the operator and passengers can make informed decisions about whether the bus should be serviced or if the air quality inside the bus is adequate. Such a vehicle monitoring would require a DT of maturity level 3 (Syncronization).
Level 6 (interoperability among digital twins) is required to improve the entire transportation system through joint decision-making between the DTs involved.

\subsection{Autonomous vehicles}

Digital twins are perceived as a means to automate the decision making process of autonomous vehicles \cite{almeaibed2021digital}. 

Recent research emphasizes that a digital twin for autonomous vehicles, capable of autonomous driving, must meet four criteria, as shown in Table \ref{Table:vehicle_levels}. However, these criteria require varying levels of maturity and can be gradually incorporated to provide value to the vehicle and its passengers before enabling full autonomous driving.
As DT-driven autonomous vehicles are significantly restricted in open environments due to legal restrictions, safety concerns and mixed traffic regulations, this section distinguishes between open and closed environments. 

\begin{table*}
    \caption{Vehicle Digital Twin Requirements (adapted from \cite{wright2020tell}) and their required level of maturity.}
    \centering
    \renewcommand{\arraystretch}{1.3}
    \begin{tabular}{ P{13.0cm} p{4.0cm}}
        \toprule
        \textbf{Required vehicle digital twin component} & \textbf{Required maturity level}\\
        \midrule
        An accurate and updatable model of a real-world driving environment, including other vehicles and pedestrians, as well as some description of weather and other atmospheric and environmental effects. & level 3 \\
        Models that simulate the response of the sensors deployed on the vehicle based on carefully chosen data from tests of these sensors. & level 3\\
        A model of the vehicle's response to driving commands (e.g., steering changes, braking, etc.) that takes road surface conditions into account. & Level 5  \\
        Autonomous vehicle control algorithms. & Level 5 \\
        \bottomrule
    \end{tabular}
    \label{Table:vehicle_levels}
\end{table*}

\subsubsection{Closed environments}
Autonomous vehicles driven by DTs are increasingly being developed and deployed in the context of closed environments, such as mining \cite{ai2023pmworld}, warehouses \cite{barosan2020development} or port terminals \cite{yang2022digital}. DTs in these environments are driven by the need for automation to increase efficiency and safety, and their environments are characterized by little or no generic traffic, where vehicles have relatively low speeds, short stopping distances, and a well-defined layout. Although autonomous driving on public roads still faces various technical and legal challenges, it is worth noting that in confined areas such as mining sites, port terminals, and distribution centers, some of these restrictions do not apply \cite{barosan2020development}. 

Presenting the application of DT-controlled autonomous vehicles in mining, the authors state in paper \cite{hazrathosseini2023advent} that the use of DT-controlled autonomous vehicles in mining is not an option, but rather a necessity. Real-life case studies of digital twins implemented in several mining companies indicate significant improvements in fleet production, over speeding incidents, and cycle delays. Specifically, a 33\% increase in fleet production, a 43\% reduction in over speeding incidents, and a 25\% decrease in cycle delays were achieved \cite{hazrathosseini2023advent}.
In paper \cite{ai2023pmworld}, the authors propose a digital twin-based virtual-real interactive system for autonomous mining vehicles. The system includes descriptive, predictive, and prescriptive intelligence, which can efficiently promote safety, sustainability, and smartness in mining.  

DTs have been applied to test and simulate various scenarios involving docking, maneuvering, and parking in the distributed central areas of warehouses, enabling autonomous vehicle operations. In paper \cite{barosan2020development}, the authors propose a DT-driven Virtual Simulation Environment (VSE) that allows for the integration of multiple trucks, enabling the development and testing of applications using platooning and multi-truck driving scenarios. In addition, the VSE enables the definition of different truck and trailer combinations using the vehicle setup feature, allowing for the testing of various usage scenarios.

DTs are also utilized in the port context to enable the operation of autonomous vehicles. In paper \cite{yang2022digital}, the authors propose a DT that facilitates autonomous truck operations within the port terminal. The proposed system is based on the cooperation of 5G mobile edge computing, real-time kinematic, and 5G base stations.
In paper \cite{widyotriatmo2024leveraging}, the authors utilize a DT to generate a virtual model of the container truck and its environment, allowing the autonomous docking control system to be tested and refined prior to implementation.

\subsubsection{Open environments}
The development of autonomous vehicles for use in open environments, such as cities, has become a significant technological challenge in recent years. This is due to the need to address multiple, often conflicting concerns such as safety, comfort, efficiency, and effectiveness \cite{thonhofer2023infrastructure}. In this context, digital twins are recognized as an enabler tool to optimize the efficiency and reliability of electric (autonomous) vehicles \cite{van2021beyond}. In paper \cite{almeaibed2021digital}, the authors investigate the potential of digital twin technology to automate decision-making processes within automated vehicles using radar sensor data. The paper places particular emphasis on safety and security aspects. 

In paper \cite{thonhofer2023infrastructure}, the authors propose a DT-driven decision support platform in the context of connected, automated mobility and intelligent road service. 

In paper \cite{samak2023autodrive}, the authors proposes AutoDRIVE, a DT-driven platform that integrates real and virtual worlds into a common toolchain. It is suitable for prototyping and validating autonomy solutions for various use cases, including autonomous parking, driver behavioral cloning, intersection traversal, and smart city management.

Table \ref{Table:levels_of_autonomy} provides an overview about the different autonomy levels for autonomous driving. Once control of the vehicle has been achieved through softwareization and subsequent computational optimization of its processes, a level 5 digital twin is required. However, to reach its full potential, level 6 is required to enable joint decision making among multiple digital twins and thus enable fleet decisions. 

\begin{table}
    \caption{Levels of autonomy in transportation. Adapted from \cite{almeaibed2021digital}.}
    \centering
    \renewcommand{\arraystretch}{1.3}
    \begin{tabular}{ P{0.8cm} p{7.0cm}}
        \toprule
        \textbf{Level} & \textbf{Description of the level of autonomy}\\
        \midrule
        0 & No driving automation \\
        1 & The driver has full control of the vehicle's main functions. \\
        2 & The driver is responsible for the safe operation of the vehicle, while certain control functions are automated. \\
        3 & The driver remains responsible for safely operating the vehicle and monitoring the roadway, while some primary control functions are automated.\\
        4 & Autonomous driving in specific environmental and traffic situations.\\
        5 & Full self-driving capabilities without any human intervention and without environmental or traffic restrictions.\\
        \bottomrule
    \end{tabular}
    \label{Table:levels_of_autonomy}
\end{table}

Another area that offers great potential for DT-driven vehicle optimization is last mile distribution (LMD). LMD poses significant challenges for businesses, including high costs, low efficiency, and environmental concerns. LMD constitutes up to 50\% of total delivery costs and improving LMD efficiency can yield substantial savings for companies \cite{sorooshian2022toward, vanelslander2013commonly}. Businesses are actively exploring novel technologies to enhance value creation in LMD. Among these are robots and autonomous vehicles, such as driverless sidewalk delivery robot \cite{simoni2020optimization, hoffmann2018regulatory}. These innovations need to be at least level 5 (see Figure \ref{Figure:maturity levels}) to hold promise for a more sustainable and efficient LMD system, when this is achieved truck milage may be reduced by 60 \% according to one delivery robot study \cite{ostermeier2022cost}.

To better monitor and optimize vehicle movements at the city level, cities can use the Mobility Data Specification (MDS) framework. MDS is a standardized framework for communication and data exchange between urban municipalities and private transport service providers, including e-scooter and bike-share companies. Through MDS, cities can efficiently share and validate policy information digitally, facilitating effective vehicle management and leading to improved outcomes for residents \cite{MDS}.

\section{Concluding discussion}
Like many other fields, vehicles also benefit from the increasing maturity of digital twins. One challenge is to assess how mature the vehicle digital twin needs to be to achieve specific goals, ranging from vehicle condition monitoring to autonomous driving. This paper fills this gap by presenting the latest developments in DT-driven research for freight, passenger, and autonomous vehicles, and discusses what level of digital twin maturity is required to enable their respective objectives.
It is concluded that digital twins are already being used to monitor critical vehicle components, and initial test beds for performing DT-enabled autonomous vehicle operations in closed environments exist. However, there are still several challenges that need to be addressed, such as standardization, interoperability, safety, and security, before digital twins can reach their full potential.

The conducted literature review on recent advances in DT-driven vehicle efficiency gains reveals that the application of DTs to vehicles is no longer limited to the design and development phase, but is increasingly applied in the operational phase. 
Compared to other complex applications such as ports and cities, vehicles have the potential to receive and exchange data and knowledge from other vehicle instances and their respective DT instances that correspond to the same prototype. This allows for cross-vehicle improvements to be achieved within the framework of the DT aggregate concept.
This leads to an increasing softwareisation of vehicles, resulting in DT-driven applications such as real-time data-driven maintenance inspections, semi-autonomous vehicle operations, and joint optimisation of multiple vehicle operations using vehicle-to-vehicle communication, such as in the case of platooning or fully autonomous operations.

The analyses of this paper indicate that digital twins are seen as an enabler for improving the efficiency of different types of vehicles, although their purposes differ, such as prioritizing the comfort and well-being of passengers in public transport or providing low-cost and sustainable disruption-free freight transport. 

Many of the currently used digital twins are designed for monitoring critical vehicle parameters. For example, they can track the state of health of the battery, which is vital for correcting anomalies, monitoring the remaining lifespan of critical vehicle components, or tracking air quality and passenger conditions. In this case, they often act as a digital shadow, reaching a maturity level of stage 3 (Synchronization). However, there are also cases where digital twins already take direct control over some vehicle functions, especially in the case of safety techniques, and thus reach a digital twin level of 5 (Automation). 

One of the key visions in the field of automotive engineering regarding the use of the digital twin is to accelerate and enable autonomous driving, which would require digital twins of level 6 (interoperability) to enable the respective digital twins of the vehicles and the environment to perform joint decision-making. Although there are promising pilot projects in closed environments, implementation in public spaces is still a major challenge. This is due to the lower level of complexity of closed environments, which results in lower security, standardization, and interoperability requirements. A comprehensive discussion of barriers and potential pathways towards enabling interoperability among digital twins leading to joint decision-making, is discussed in paper \cite{klar2023digital}.

Recent research to address safety, comfort, efficiency, and regulatory compliance in autonomous driving suggests that approaches consistent with these efforts include information sharing, connected driving, and infrastructure supported driving (infrastructure provides information to the vehicle). However, there are still many unresolved questions, such as what information is best to share, in what form and quality, to ensure safety and security, and ensure trust \cite{thonhofer2023infrastructure}.

Another key issue in open space is security, as self-driving vehicles are vulnerable to attacks since they must communicate with other vehicles and external networked software/hardware infrastructure. An overview of potential attacks and solutions is presented in paper \cite{almeaibed2021digital}.

\section{Acknowledgement}

This work has been supported by Trafikverket Sweden as part of the Triple F (MODIG-TEK) project under Grant 2019.2.2.16, and from the European Union’s Horizon Europe programme under grant agreement No101057779 (project TWINAIR).

\ifCLASSOPTIONcaptionsoff
  \newpage
\fi



\bibliographystyle{IEEEtran}
\bibliography{bibtex/bib/references.bib}
%



%

\newpage
\begin{IEEEbiography}{Robert Klar}
Biography text here.
\end{IEEEbiography}

\begin{IEEEbiography}{Vangelis Angelakis}
Biography text here.
\end{IEEEbiography}







\end{document}